\documentclass[aps,prl,reprint,notitlepage,showpacs,nofootinbib,superscriptaddress,twocolumn%,raggedbottom
,nofloats,preprintnumbers]{revtex4-1}
\usepackage{lineno}
\usepackage{graphicx}
\usepackage[utf8]{inputenc} 
\usepackage{amsmath}
\usepackage{amssymb}
\usepackage{ulem}
\usepackage{float}
\usepackage{comment}
\usepackage{slashed}

\usepackage[T1]{fontenc} % if needed
\usepackage{parskip}
\usepackage{bm}         % for bold math characters in a math environment
\usepackage{array}
\usepackage{multirow}   % analogue to \multicolumn
\usepackage{booktabs}   % for fancy looking tables

\usepackage{textcomp}
\usepackage{amssymb}

\usepackage{enumerate}
\usepackage{caption}
\usepackage{subcaption}
\usepackage{ragged2e}
 \DeclareCaptionJustification{justified}{\justifying}
\usepackage{amsmath}
\usepackage{amssymb}
\usepackage{indentfirst}
\usepackage[usenames,dvipsnames]{xcolor}
\usepackage[colorlinks=true,citecolor=darkred,urlcolor=darkred, pdfborder={0 0 0}]{hyperref}
\definecolor{darkred}{rgb}{0.6,0,0}

\definecolor{linkcolor}{rgb}{0,0,0.5}
\newcommand {\ignore}[1]{}

\definecolor{bostonuniversityred}{rgb}{0.8, 0.0, 0.0}

\def\gsim{\raise0.3ex\hbox{$\;>$\kern-0.75em\raise-1.1ex\hbox{$\sim\;$}}}
\def\lsim{\raise0.3ex\hbox{$\;<$\kern-0.75em\raise-1.1ex\hbox{$\sim\;$}}}

\usepackage{soul}
\definecolor{mightnightblue}{RGB}{25,25,112}

\definecolor{brown}{rgb}{0.59, 0.29, 0.0}

\def\21{$\mathrm{SU(2)_L \otimes U(1)_Y}$}

\newcommand{\AddrUNAM}{ {\it Instituto de F\'{\i}sica, Universidad Nacional Aut\'onoma de M\'exico, A.P. 20-364, Ciudad de M\'exico 01000, M\'exico.}}

\DeclareMathOperator\erf{erf}
\begin{document}

%\runningpagewiselinenumbers
%\lipsum[1-20]
%\linenumbers

\title{\boldmath\color{BrickRed} Physics reach of a low threshold scintillating argon bubble chamber in  coherent elastic neutrino-nucleus scattering reactor experiments}

\author{L. J. Flores}\email{luisjf89@fisica.unam.mx }
\affiliation{\AddrUNAM}

\author{Eduardo Peinado}\email{epeinado@fisica.unam.mx}
\affiliation{\AddrUNAM}

\collaboration{CE$\nu$NS Theory Group at IF-UNAM}

\author{E.~Alfonso-Pita}\email{ernestoalfonso@estudiantes.fisica.unam.mx }
\affiliation{\AddrUNAM}

\author{K.~Allen}
\affiliation{Department of Physics, Indiana University South Bend, South Bend, Indiana 46634, USA}

\author{M.~Baker}
\affiliation{Department of Physics, University of Alberta, Edmonton, T6G 2E1, Canada}

\author{E.~Behnke}
\affiliation{Department of Physics, Indiana University South Bend, South Bend, Indiana 46634, USA}

\author{M.~Bressler}
\affiliation{Department of Physics, Drexel University, Philadelphia, Pennsylvania 19104, USA}

\author{K.~Clark}
\affiliation{Department of Physics, Queen's University, Kingston, K7L 3N6, Canada}

\author{R.~Coppejans}
\affiliation{Department of Physics and Astronomy, Northwestern University, Evanston, Illinois 60208, USA}
\affiliation{Colegio de F\'{\i}sica Fundamental e Interdiciplinaria de las Am\'ericas (COFI), San Juan, Puerto Rico 00901, USA}

\author{C.~Cripe}
\affiliation{Department of Physics, Indiana University South Bend, South Bend, Indiana 46634, USA}

\author{M.~Crisler}
\affiliation{Fermi National Accelerator Laboratory, Batavia, Illinois 60510, USA}

\author{C.~E.~Dahl}
\affiliation{Department of Physics and Astronomy, Northwestern University, Evanston, Illinois 60208, USA}
\affiliation{Fermi National Accelerator Laboratory, Batavia, Illinois 60510, USA}

\author{A.~de~St.~Croix}
\affiliation{Department of Physics, Queen's University, Kingston, K7L 3N6, Canada}

\author{D.~Durnford}
\affiliation{Department of Physics, University of Alberta, Edmonton, T6G 2E1, Canada}

\author{P.~Giampa}
\affiliation{SNOLAB, Lively, Ontario, P3Y 1N2, Canada}

\author{O.~Harris}
\affiliation{Northeastern Illinois University, Chicago, Illinois 60625, USA}

\author{P.~Hatch}
\affiliation{Department of Physics, Queen's University, Kingston, K7L 3N6, Canada}

\author{H.~Hawley-Herrera}
\affiliation{Department of Physics, Queen's University, Kingston, K7L 3N6, Canada}

\author{C.~M.~Jackson}
\affiliation{Pacific Northwest National Laboratory, Richland, Washington 99354, USA}

\author{Y.~Ko}
\affiliation{Department of Physics, University of Alberta, Edmonton, T6G 2E1, Canada}

\author{C.~B.~Krauss}
\affiliation{Department of Physics, University of Alberta, Edmonton, T6G 2E1, Canada}

\author{N.~Lamb}
\affiliation{Department of Physics, Drexel University, Philadelphia, Pennsylvania 19104, USA}

\author{M.~Laurin}
\affiliation{Département de Physique, Université de Montréal, Montréal, H3T 1J4, Canada}

\author{I.~Levine}
\affiliation{Department of Physics, Indiana University South Bend, South Bend, Indiana 46634, USA}

\author{W.~H.~Lippincott}
\affiliation{Department of Physics, University of California Santa Barbara, Santa Barbara, California 93106, USA}

\author{R.~Neilson}
\affiliation{Department of Physics, Drexel University, Philadelphia, Pennsylvania 19104, USA}

\author{S.~Pal}
\affiliation{Department of Physics, University of Alberta, Edmonton, T6G 2E1, Canada}

\author{M.-C.~Piro}
\affiliation{Department of Physics, University of Alberta, Edmonton, T6G 2E1, Canada}

\author{Z.~Sheng}
\affiliation{Department of Physics and Astronomy, Northwestern University, Evanston, Illinois 60208, USA}

\author{E.~V\'azquez-J\'auregui}\email{ericvj@fisica.unam.mx }
\affiliation{\AddrUNAM}
\affiliation{Department of Physics, Laurentian University, Sudbury, P3E 2C6, Canada}

\author{T.~J.~Whitis}
\affiliation{Department of Physics, University of California Santa Barbara, Santa Barbara, California 93106, USA}

\author{S.~Windle}
\affiliation{Department of Physics, Drexel University, Philadelphia, Pennsylvania 19104, USA}

\author{R.~Zhang}
\affiliation{Department of Physics, University of California Santa Barbara, Santa Barbara, California 93106, USA}

\author{A.~Zu\~niga-Reyes}
\affiliation{\AddrUNAM}

\collaboration{SBC Collaboration}

\vspace{0.7cm}

\preprint{FERMILAB-PUB-21-016-AE-E-LDRD-ND}

\begin{abstract}
\vspace{0.3cm}
\noindent The physics reach of a low threshold (100~eV) scintillating argon bubble chamber sensitive to Coherent Elastic neutrino-Nucleus Scattering (CE$\nu$NS) from reactor neutrinos is studied. The sensitivity to the weak mixing angle, neutrino magnetic moment, and a light $Z'$ gauge boson mediator are analyzed. A Monte Carlo simulation of the backgrounds is performed to assess their contribution to the signal.
The analysis shows that world-leading sensitivities are achieved with a one-year exposure for a 10~kg chamber at 3~m from a 1~MW$_{th}$ research reactor or a 100~kg chamber at 30~m from a 2000~MW$_{th}$ power reactor. Such a detector has the potential to become the leading technology to study CE$\nu$NS using nuclear reactors.

\end{abstract}

\keywords{Coherent Elastic neutrino-Nucleus Scattering, bubble chambers, extra gauge boson, weak mixing angle, neutrino magnetic moment}
\maketitle

\section{Introduction}

The detection of neutrinos produced at nuclear reactors via Coherent Elastic neutrino-Nucleus Scattering (CE$\nu$NS) presents both an experimental challenge and a host of new opportunities in neutrino physics.  Measurements of CE$\nu$NS to date have relied on pion decay-at-rest neutrino sources \cite{Akimov:2017ade,Akimov:2020pdx}, measuring $O(10)$-keV nuclear recoils and taking advantage of the $\sim$10$^{-4}$ duty cycle of the Spallation Neutron Source at Oak Ridge National Laboratory.  By contrast, the few-MeV neutrinos produced by nuclear reactors give a continuous rate of sub-keV nuclear recoils, requiring an order-of-magnitude reduction in threshold and many-order-of-magnitude reduction in backgrounds.  The payoff, if these challenges are met, includes precision measurements of neutrino properties enabled by the up to $\times10^5$-higher neutrino flux, fully coherent scattering of low-energy neutrinos and pure anti-electron neutrino flavor. A variety of detector technologies are now in an experimental race to make the first reactor CE$\nu$NS observation~\cite{Abreu:2020bzt,Hakenmuller:2019ecb,Belov:2015ufh,Agnolet:2016zir,Billard:2016giu,Moroni:2014wia,Akimov:2017hee,Fernandez-Moroni:2020yyl,Choi:2020gkm,Wong:2016lmb,Strauss:2017cuu}.\\
\indent This paper explores the potential neutrino physics reach of a new enabling technology for reactor CE$\nu$NS detection, the liquid-noble (scintillating) bubble chamber.  As in dark matter direct detection, this technique achieves the necessary background reduction by distinguishing between nuclear recoils (signal) and electron recoils (backgrounds from $\gamma$-rays and beta decays), but where existing detection techniques lose discrimination at nuclear recoil energies below $\sim$1~keV~\cite{Agnese:2016cpb,Akerib:2018lyp,PhysRevD.99.112009,PhysRevD.100.072009,PhysRevD.100.082006}, the liquid-noble bubble chamber may maintain discrimination at nuclear recoil energies as low as 100~eV.  This study takes a specific scenario motivated by the work of the Scintillating Bubble Chamber (SBC) Collaboration~\cite{Giampa_2020,Baxter_2017}, but qualitatively the results would apply to any technique that (1) has a measurable and calibrated response to 100-eV nuclear recoils, (2) eliminates electron-recoil backgrounds through discrimination, (3) is able to measure nuclear-recoil backgrounds in-situ through side-band analyses, and (4) scales to 10--100-kg target masses.\\
\indent A description of the detailed experimental scenarios considered and the ways in which they meet the above requirements is given in the next section (Experiment Description).  The following section (Physics Reach) investigates the sensitivity of these experiments to the weak mixing angle, the neutrino magnetic moment, and a Non-Standard Interaction (NSI) through a $Z'$ gauge boson mediator.  We conclude that reactor CE$\nu$NS provides both a realistic and powerful opportunity to constrain and discover neutrino physics beyond the Standard Model (SM).

\section{Experiment description}

\begin{figure}[t]
    \centering
    \includegraphics[width=\linewidth]{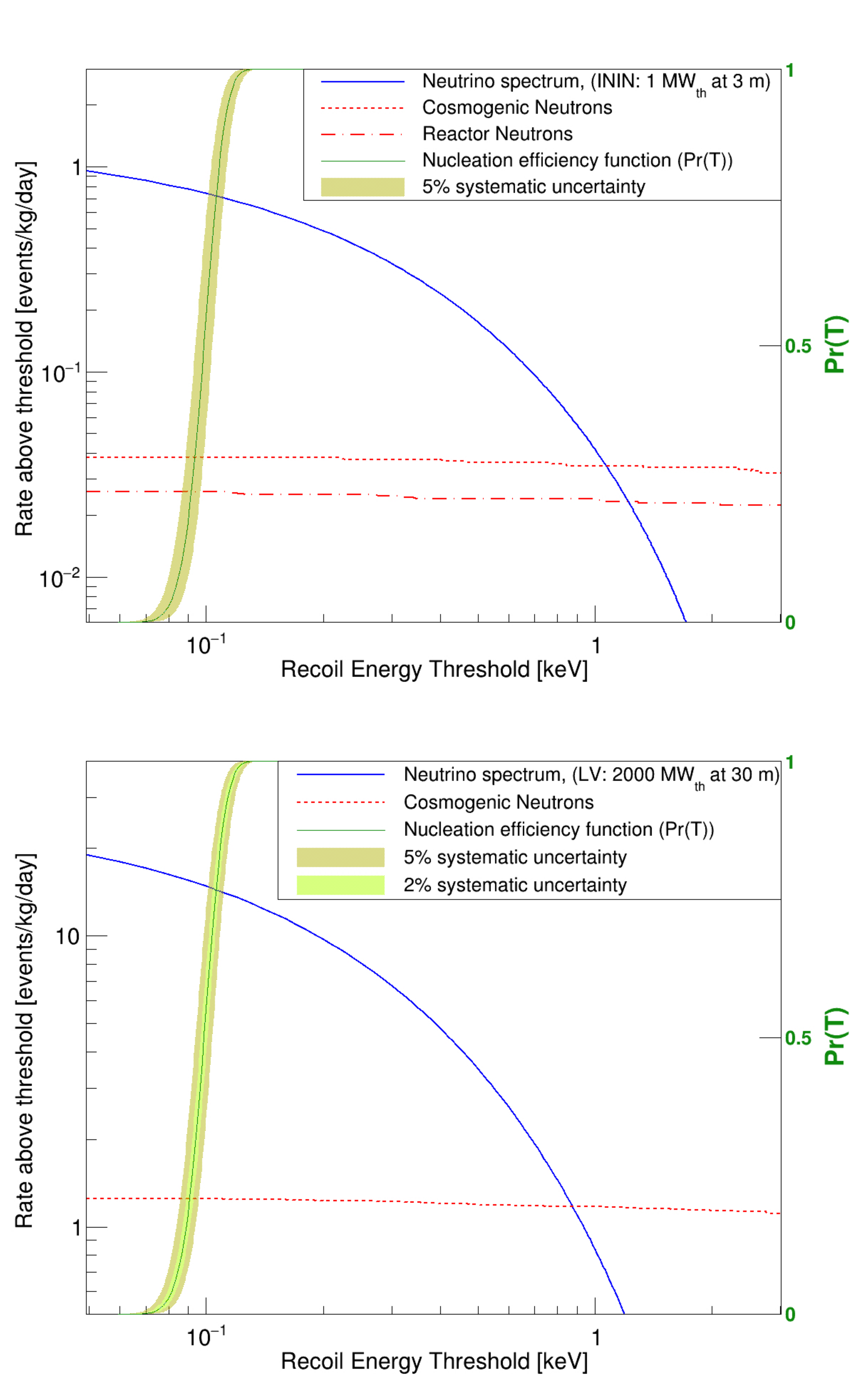}
    \caption{Signal and neutron background rates above threshold for setups A (top) and B (bottom). Backgrounds in setup A come from reactor and cosmogenic neutrons while only cosmogenic neutrons are shown in setup B since at 30~m (usually outside of the reactor building) the backgrounds produced from the core are negligible. The assumed Normal Cumulative Distribution Function (Gaussian CDF) to describe the nucleation efficiency and its systematic uncertainty, as described in the Calibration section, are are also presented (green).}
    \label{fig:neutrinorate}
\end{figure}

Superheated liquids have been used for over a decade by dark matter direct detection experiments searching for Weakly Interacting Massive Particles (WIMPs), most recently in the PICO Collaboration's fluorocarbon bubble chambers \cite{Amole:2015lsj,Amole:2015pla,Amole:2019fdf,PhysRevLett.118.251301}. Nuclear recoils in the superheated targets of these devices create a single bubble, which, if the nuclear recoil energy is above a threshold set by the temperature and pressure of the target fluid, grows within a few milliseconds to macroscopic size\footnote{This process is described by the Seitz model of bubble nucleation~\cite{Seitz_1958}.}. These detectors are completely insensitive to electron recoils (nucleation efficiency~$<$10$^{-10}$ in C$_3$F$_8$ and CF$_3$I) when operated with nuclear recoil thresholds above a few keV \cite{PhysRevD.100.082006}, since the bubble nucleation depends not only on the energy deposited by the incoming particle but also on its stopping power.\\
\indent Work by the SBC Collaboration has shown that liquid-noble bubble chambers are able to operate at much higher degrees of superheat (lower thresholds) than fluorocarbon-based detectors~\cite{Baxter_2017}.  Most recently, a xenon bubble chamber was operated at thresholds down to 500~eV\footnote{Low-threshold performance from private communication, publication in preparation.} while remaining insensitive to electron recoil backgrounds, proving the feasibility of reducing the threshold with noble liquids and demonstrating simultaneous bubble nucleation and scintillation by nuclear recoils. The SBC Collaboration is currently designing and building a 10-kg liquid argon (LAr) bubble chamber with a target energy threshold of 100~eV. The higher superheat necessitates higher pressure when recompressing the fluid following bubble nucleation than was possible in the xenon bubble chamber. Previous LAr bubble chambers have operated with even lower thresholds~\cite{Berset:1982,Harigel:1981}, but for the current chamber, success is critically dependent on electron-recoil rejection being retained at better than the $10^{-8}$ level at this low threshold, which must be experimentally verified. This detector will be equipped with Silicon Photomultipliers (SiPMs) to collect scintillation light generated in the target fluid, used to veto high-energy events mostly produced by cosmogenic and muon-induced neutrons, as well as reactor neutrons (from the core and through ($\gamma$, n) reactions in the materials). Due to the light collection efficiency of the SiPM system, no detected scintillation light is expected for reactor CE$\nu$NS interactions and other low-energy recoils ($\lessapprox5$-keV nuclear recoil equivalent). These experimental techniques and developments open a new window of opportunity to study CE$\nu$NS in nuclear reactors using noble liquids operated at very low thresholds and free of electron recoil backgrounds.\\
\indent Two main detector configurations are considered in this work: a 10-kg LAr chamber operated at a 100-eV energy threshold and located 3~m from a 1-MW$_{th}$ reactor (setup A)\footnote{A TRIGA Mark III research reactor located at the National Institute for Nuclear Research (ININ) near Mexico City is being explored as a possible location.}, where ${\sim} 8$ neutrino events/day above threshold are expected; and a 100-kg LAr chamber operated at the same threshold and located 30~m from a 2000-MW$_{th}$ power reactor (setup B)\footnote{The Laguna Verde (LV) power reactor consisting of two BRW-5 (Boiling Water Reactors) units located in the east coast of Mexico in the Gulf of Mexico is also explored as a possible location.}, where ${\sim}1570$ neutrino events/day above threshold are expected. Fig.~\ref{fig:neutrinorate} shows the signal and neutron background rates above threshold for the setups described. These configurations assume a 2.4\% uncertainty in the anti-neutrino flux (as per the uncertainty in the prediction of the Huber$+$Mueller model~\cite{Huber:2011wv,Mueller:2011nm}) and 5\% systematic uncertainty in the energy threshold. A third configuration named setup B(1.5) is also considered, with the same parameters as setup B but with a 1.5\% uncertainty in the anti-neutrino flux (as per the uncertainty in the Daya Bay measurement from their reactors~\cite{Adey:2018qct}) and a 2\% systematic uncertainty in the energy threshold. The parameters for this third configuration are considered aggressive and are reported to present the maximum physics reach that could be achieved. Table~\ref{table:setups} summarizes the setups considered and the relevant parameters assumed. The physics reach reported in this work is not specific to the locations described; they are just sites currently explored.
\begin{table}[htb]
\centering
\begin{tabular}{|c|c|c|c|c|c|c|c|}
\hline
Setup & LAr  & Power      & Distance & Anti-$\nu$ flux &  Threshold    \\
      & mass &            &          & uncertainty     &  uncertainty  \\
      & (kg) & (MW$_{th}$) & (m)      & (\%)            &  (\%)         \\
\hline
A      & 10  & 1    & 3  & 2.4 & 5 \\
B      & 100 & 2000 & 30 & 2.4 & 5 \\
B(1.5) & 100 & 2000 & 30 & 1.5 & 2 \\
\hline
\end{tabular}
\caption{Relevant parameters assumed for the setups considered. Threshold uncertainty is the estimated uncertainty in the nuclear recoil energy threshold.}
\label{table:setups}
\end{table}

\subsection{Backgrounds}

A GEANT4 \cite{Geant4_0,Geant4_1,Geant4_2} Monte Carlo simulation was developed to estimate the main background contributions, primarily neutrons from cosmic rays and the reactor itself. While backgrounds from cosmic rays can be statistically subtracted with a reactor-off dataset, reactor-induced backgrounds must be estimated with {\it in-situ} measurements and simulations. Backgrounds were studied in the explored sites at the National Institute for Nuclear Research (ININ) near Mexico City, for the 1 MW$_{th}$ reactor configuration (cosmic muon rate of $146\,\mu/m^2/s$~\cite{4437209}), and at Laguna Verde (only from cosmic rays and not from the reactor) on the east coast of Mexico, for the 2000 MW$_{th}$ reactor (cosmic muon rate of $104\,\mu/m^2/s$~\cite{4437209}).

\subsubsection{Setup A}

\begin{figure}[t]
    \centering
    \includegraphics[width=\linewidth]{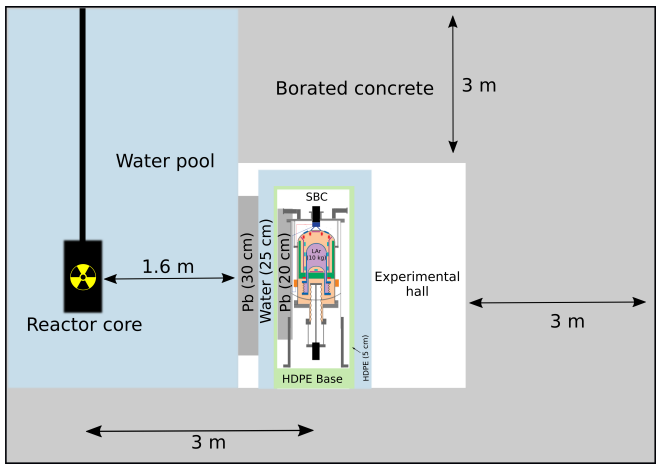}
    \caption{Conceptual design of the configuration for setup A at the ININ experimental hall. The distance between the reactor core center and the bubble chamber is 3~m. Accuracy of a few mm is expected based on survey work developed at ININ.}
    \label{fig:ININhall}
\end{figure}

For setup A, the model includes the experimental hall at ININ, which is surrounded by approximately 3~m of high-density borated concrete that will act as a shield for cosmogenic neutrons. Moreover, the shielding model features 25~cm of water and 5~cm of polyethylene surrounding the detector, a 30 cm thick Pb-wall between the water pool and the shielding, and another 20 cm thick Pb-wall next to the bubble chamber. The distance between the reactor core center and the bubble chamber is 3~m, including 1.6~m of water shielding provided by the reactor pool. The shielding configuration for this setup is illustrated in Fig.~\ref{fig:ININhall}.\\
\indent Neutrons produced by the reactor core are estimated using a measurement at ININ performed as part of the radiation programme \cite{doi:10.1063/1.4813460,doi:10.1063/1.4927182}. Nuclear recoils produced by ($\gamma$, n) reactions and Thomson ($\gamma$-nucleus elastic) scattering~\cite{Robinson:2016imi} from $\gamma$-rays produced by the reactor are estimated using a gamma flux simulation for a TRIGA Mark III reactor, obtained with an MCNP model of the core~\cite{Agnolet:2016zir}.\\ 
\indent Cosmogenic neutrons are estimated with a simulation of the neutron flux using the code CRY~\cite{4437209} and neutrons induced by muons interacting with materials in the deployment site are estimated using the parametrization from \cite{PhysRevD.73.053004} in water and concrete.\\ 
\indent The simulations predict $0.25$ events/day above threshold (3.1\% of the signal) from backgrounds produced by the reactor. Of these, $0.03$ events/day (0.4\% of the signal) are from reactor neutrons, $0.16$ events/day (2.0\% of the signal) are from $^{2}$H($\gamma$,n) reactions in the water, and $0.06$ events/day (0.7\% of the signal) are from $^{208}$Pb($\gamma$,n) and $^{207}$Pb($\gamma$,n). The shielding concept proposed reduces the gamma flux from the reactor core to ${\sim}1$ Hz in the LAr target volume. At this rate electron recoil backgrounds are negligible given the expected insensitivity to these events. Thomson scattering is expected to contribute $0.0002$ events/day ($<$0.01\% of the signal).\\
\indent The simulations also predict $0.85$ events/day above threshold (10.8\% of the signal) from backgrounds produced by cosmic rays, including $0.38$ events/day from cosmogenic neutrons (4.8\% of the signal) and $0.47$ events/day (6.0\% of the signal) from muon-induced neutrons in water and concrete.

\subsubsection{Setups B and B(1.5)}

\indent For setups B and B(1.5), only simulations for cosmogenic and muon-induced neutrons were considered, since at 30~m (usually outside of the reactor building) the backgrounds produced from the core are negligible. Shielding consisting of 3~m of water and 50~cm of polyethylene is included in this simulation, which reduces the backgrounds from cosmic rays to 180 events/day above threshold (11.5\% of the signal), including 125 events/day from cosmogenic neutrons and 55 events/day from muon-induced neutrons in the water shield.\\

\indent Backgrounds from internal radioactivity are negligible for all configurations, accounting for approximately $0.003$ events/day above threshold (<1\% of the signal), where the purity of the components assumed is similar to the materials used in bubble chambers built by the PICO Collaboration~ \cite{Amole:2015lsj,Amole:2019fdf}.\\
\indent Overall, the background contribution to the signal is estimated to be on the order of 5\% (from the reactor) and 11\% (from cosmic rays) for setup A, and 12\% (from cosmic rays) for setups B and B(1.5). The physics reach reported in this manuscript assumes these background levels, which do not consider the ability to veto $\gtrapprox5$-keV recoils by their scintillation light. A conservative systematic uncertainty of 10\% is assumed for reactor backgrounds, larger than the 5\% uncertainty in the ININ neutron flux measurements. Reactor backgrounds can also be characterized {\it in-situ} from non-signal regions (multiply-scattering neutron events and bubbles coincident with scintillation signals). Backgrounds from cosmic rays are statistically subtracted with no systematic uncertainty.

\subsection{Calibration}

The response of a bubble chamber to nuclear recoils is described by a nucleation efficiency function, representing the probability of a recoil with energy $T$ to nucleate a bubble, rising from 0 to 100\% in the vicinity of an energy threshold $E_T$. For the physics reach reported here, a Normal Cumulative Distribution Function (Gaussian CDF) is assumed, 
\begin{equation}
    Pr(T) = \frac{1}{2}\left(1 + \erf\left(\frac{T-E_T}{\sigma\sqrt{2}}\right)\right),
\end{equation}
where $E_T$ is set to 100~eV and the width $\sigma$ is set to 10~eV, shown in Fig.~\ref{fig:neutrinorate}. This functional form, and the relative sharpness of the turn-on, are chosen to approximate the shape of the observed efficiencies measured in C$_3$F$_8$~\cite{Amole:2019fdf}; the exact shape will need to be experimentally measured. A 5\% (setups A and B) or 2\% (setup B(1.5)) systematic uncertainty in $E_T$ is assumed, intended to encompass both threshold and general shape uncertainties following a calibration program. \\
\indent Low energy, nearly mono-energetic, neutrons can be produced by ($\gamma$,n) reactions in beryllium. Three photo-neutron sources, each producing different recoil energy spectra in the detector, are proposed to calibrate low-energy nuclear recoils. $^{207}$Bi-Be (94~keV neutrons), $^{124}$Sb-Be (23 and 380~keV neutrons) and $^{58}$Co-Be (9~keV neutrons) sources were simulated in the GEANT4 geometry developed for the 10-kg chamber. The simulations indicate that with sources of 1 to 100 $\mu$Ci activities, high-statistics recoil energy spectra below 8~keV, 3~keV, and 1~keV can be achieved with the $^{207}$Bi-Be, $^{124}$Sb-Be, and $^{58}$Co-Be sources, respectively. These sources would allow constraint of the nucleation efficiency function for different thermodynamic conditions. A similar technique has previously been implemented by the PICO Collaboration~\cite{Amole:2019fdf}. \\
\indent Blindness to electron recoils allows for a novel additional calibration with nuclear recoils from Thomson scattering. For example, 1.33, 1.41 and 1.46~MeV $\gamma$-rays from $^{60}$Co, $^{152}$Eu and $^{40}$K produce nuclear recoil spectra with sharp cut-offs at 95, 107 and 115~eV respectively, and would provide strong constraints on the nucleation efficiency for recoils $\sim$100~eV. Finally, a tagged recoil calibration may be possible with thermal neutrons. De-excitation $\gamma$-rays from neutron capture on $^{40}$Ar result in a recoiling $^{41}$Ar nucleus with energy peaked $\sim$320~eV.

\section{Physics reach}

The physics reach of the setups described above is investigated for a one-year exposure. The SM cross-section for CE$\nu$NS, after neglecting the axial contribution, is:
\begin{equation}
\frac{d\sigma}{dT} = \frac{G_F^2}{2\pi}M_N Q_w^2 \left(2 - \frac{M_N T}{E_\nu^2}\right)F^2(q^2),
\label{eq:crossSec}
\end{equation}
where $T$ is the nuclear recoil energy, $E_\nu$ the incoming neutrino energy, $F(q^2)$ the nuclear form factor, $Q_w = Z g_p^V + N g_n^V$ is the weak nuclear charge and $M_N$, $Z$, $N$ are the nuclear mass, proton, and neutron number of the detector material, respectively.
The cross-section is convolved with the reactor anti-neutrino spectrum and the detector efficiency to compute the number of events. The theoretical prediction of the Huber$+$Mueller model~\cite{Huber:2011wv,Mueller:2011nm}, which gives a 2.4\% uncertainty in the total flux, is considered for setups A and B for neutrino energies between 2 and 8~MeV (Ref.~\cite{Kopeikin:1997ve} is used for neutrinos below 2~MeV). On the other hand, the Daya Bay experiment measured the anti-neutrino flux from their reactors with an uncertainty of $1.5\%$~\cite{Adey:2018qct}. Setup B(1.5) considers this uncertainty. The results presented are not corrected by the discrepancy between the world average of the absolute flux measured and the best prediction, which is approximately 5\%~\cite{Mention:2011,DayaBay:2016}. It is also worth mentioning that at reactor energies, the uncertainties in the form factors are negligible compared to the uncertainty in the anti-neutrino spectrum~\cite{Tomalak:2020zfh}.\\
\indent The sensitivity of this experiment is fitted with the following $\chi^2$ function:
\begin{eqnarray}\scriptsize
\chi^2 =\underset{\alpha,\beta,\gamma}{\min}&\left[\left(\frac{N_\mathrm{meas} - (1+\alpha)N_\mathrm{th}(X,\gamma)- (1+\beta)B_\mathrm{reac}}{\sigma_\mathrm{stat}} \right)^2\right. \nonumber\\ 
&\left. + \left(\frac{\alpha}{\sigma_\alpha}\right)^2 + \left(\frac{\beta}{\sigma_\beta}\right)^2 + \left(\frac{\gamma}{\sigma_\gamma}\right)^2\right],
\label{eq:chisq}
\end{eqnarray}
where $N_\mathrm{meas}$ is the measured number of events after subtracting the background from cosmogenic and muon-induced neutrons ($B_\mathrm{cosm}$), $N_\mathrm{th}(X,\gamma)$ is the theoretical prediction with the nuclear recoil threshold set to (1+$\gamma$)$\cdot$100~eV, $B_\mathrm{reac}$ is the background coming from the reactor,
$\sigma_\mathrm{stat} = \sqrt{N_\mathrm{meas} + (R+1)B_\mathrm{cosm}}$ is the statistical uncertainty, where $R$ is the ratio of reactor-on time to reactor-off time\footnote{Four months off time is assumed at ININ (R=3) and one month off time at LV (R=12). The two reactors at LV are expected to be off simultaneously at least a few days per year. The reactor at ININ usually operates two weeks per month.}, and $\sigma_{\alpha, \beta, \gamma}$ are the systematic uncertainties on the signal, background, and threshold, respectively. The variable $X$ refers to the parameter to be fitted (weak mixing angle, NSI parameters, or neutrino magnetic moment). The $\chi^2$ function is minimized over the nuisance parameters $\alpha$, $\beta$ and $\gamma$.
The systematic uncertainties have the values $\sigma_\alpha=0.024$, $\sigma_\beta=0.1$, and $\sigma_\gamma=0.05$ for setups A and B, coming from the uncertainty on the anti-neutrino flux, the reactor neutron background, and the energy threshold, respectively. The parameters $\beta$ and $\sigma_\beta$ are absent in setups B and B(1.5) since the reactor component of the background reaching the detector is negligible. The systematic uncertainties for setup B(1.5) are $\sigma_\alpha=0.015$ and $\sigma_\gamma=0.02$.
In the following analyses, $N_\mathrm{meas}$ is assumed to be the SM predicted signal.

\subsection{The Weak Mixing Angle}
\begin{figure}[t]
    \centering
    \includegraphics[width=\linewidth]{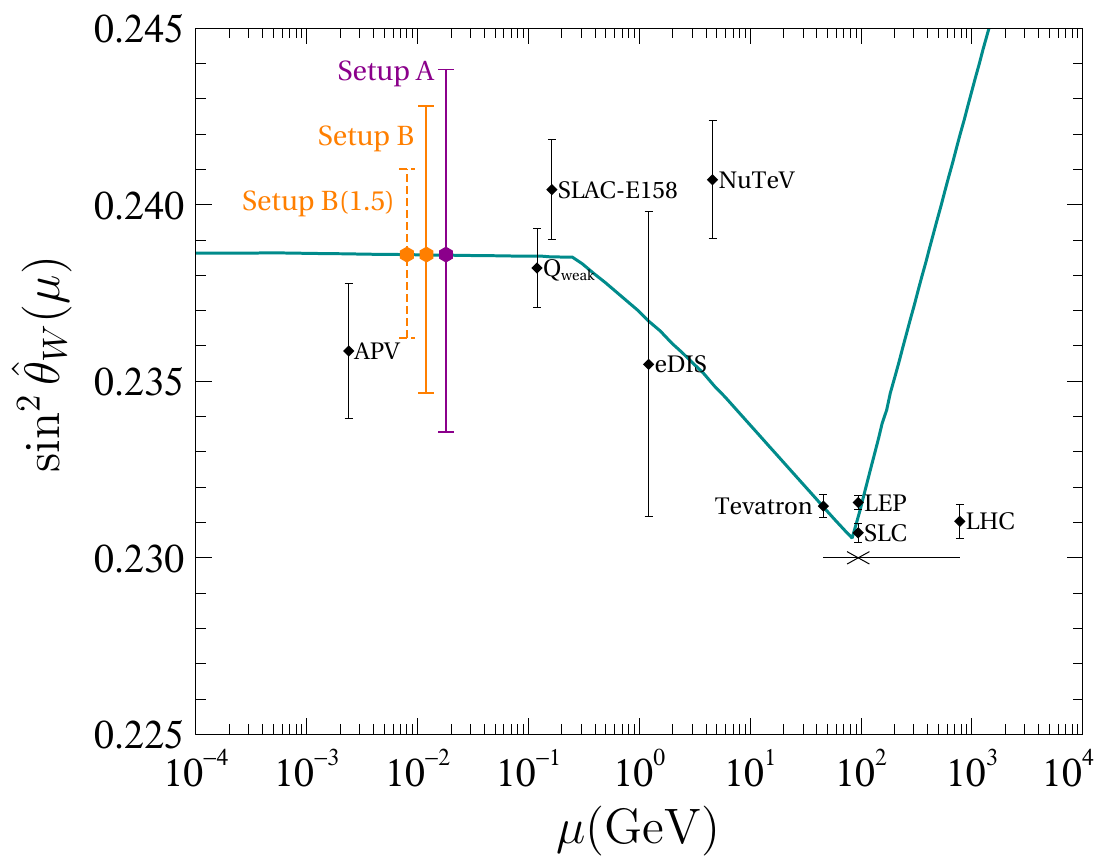}
    \caption{RGE running of the weak mixing angle in the $\overline{\mathrm{MS}}$ renormalization scheme~\cite{Erler:2017knj,Zyla:2020zbs}, as a function of the energy scale $\mu$. The expected measurements and $1\sigma$ uncertainties for setups A, B and B(1.5) are shown in solid purple, solid orange, and dashed orange, respectively.  Measurements from other experiments are also presented. Figure adapted from~\cite{Erler:2017knj}. A value of $0.209^{+0.072}_{-0.069}$ has been extracted from the CsI COHERENT data~\cite{Papoulias:2019txv} (not shown in this plot). The three setups shown in the plot have all the same value of $\mu$\,$\sim$\,$17$~MeV.}
    \label{fig:weakMixingAngle}
\end{figure}

Assuming that the experiment measures only the SM signal, a fit is performed and the value of the weak mixing angle at low energies is extracted with its corresponding uncertainty. The weak mixing angle can be extracted from the CE$\nu$NS differential cross-section through the SM weak coupling $g_p^V = 1/2 - 2 \sin^2\theta_W$. 
A fit using Eq.~\eqref{eq:chisq} is performed where $X=\sin^2\theta_W$. In Fig.~\ref{fig:weakMixingAngle} the Renormalization Group Equation (RGE) running of the weak mixing angle as a function of the energy scale is shown, in the Minimal Subtraction ($\overline{\mathrm{MS}}$) renormalization  scheme~\cite{Erler:2017knj,Zyla:2020zbs}, as well as the projections of the detectors for the setups described, and their estimated $1\sigma$ uncertainties.\\
\indent The projection obtained for the configuration assuming 1.5\% uncertainty in the reactor spectrum is not only complementary to the low-energy measurement from Atomic Parity Violation (APV)~\cite{Antypas_2018}, but is also the most sensitive among projections for several CE$\nu$NS experiments~\cite{Canas:2018rng} that assume $1.0\%$ to $1.3\%$ systematic uncertainty in the reactor spectrum. Even though there is a precise measurement from APV for the weak mixing angle at low energies, sometimes tensions in different measurements can shed light on physics beyond the Standard Model or provide a better understanding of the phenomenology.

\subsection{Light Gauge Boson Mediator}

\begin{figure}[t]
    \centering
    \includegraphics[width=\linewidth]{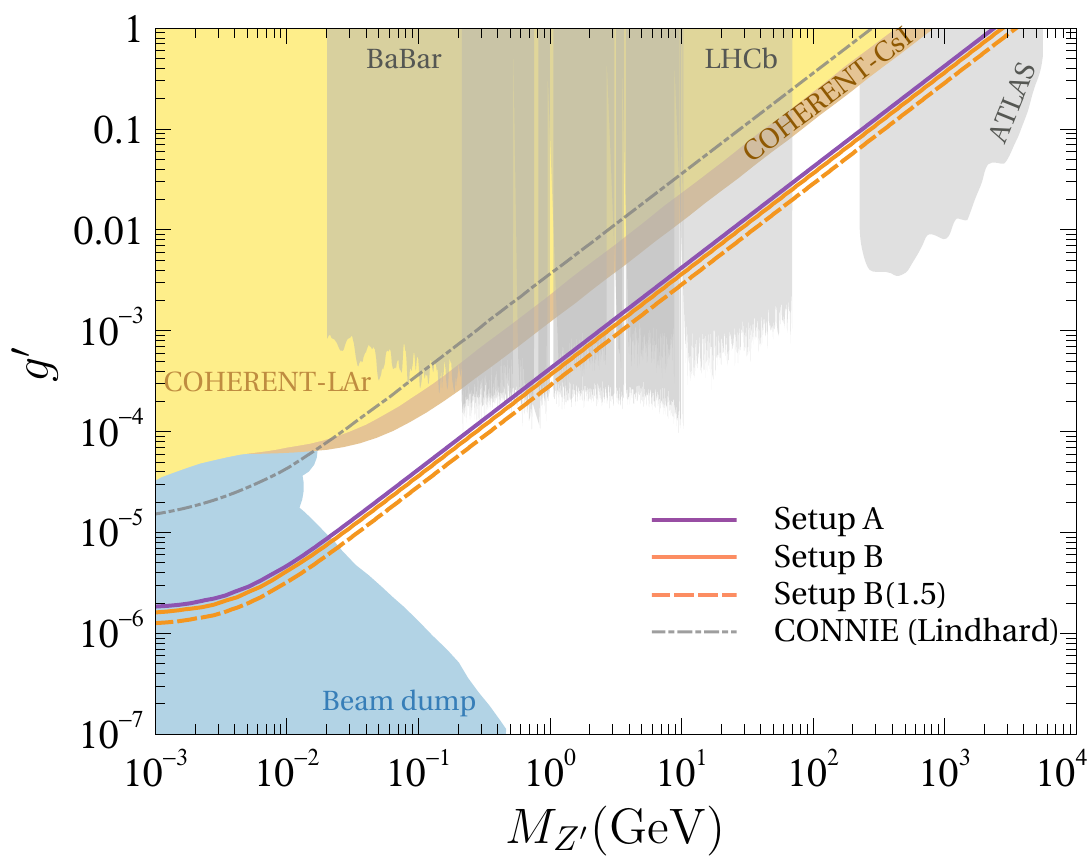}
    \caption{Exclusion limits (95\% C.L.) in the $g^\prime$-$M_{Z^\prime}$ plane. The solid purple, solid orange and dashed orange lines represent the limits for setups A, B and B(1.5), respectively. The dash-dotted gray curve is the exclusion set by CONNIE~\cite{Aguilar-Arevalo:2019zme}. The shaded brown and yellow regions correspond to the exclusions set by COHERENT, using CsI~\cite{Akimov:2017ade} and LAr~\cite{Akimov:2020pdx,Akimov:2020czh} detectors, respectively. Exclusion regions for dark photon searches from BaBar~\cite{Lees:2014xha} and  LHCb~\cite{Aaij:2019bvg} are shown in light gray, and from beam dump experiments~\cite{Bergsma:1985qz,Tsai:2019mtm,Bjorken:1988as,Riordan:1987aw,Bross:1989mp,Konaka:1986cb,Banerjee:2019hmi,Astier:2001ck,Davier:1989wz,Bernardi:1985ny} are shown in blue. These limits were obtained in the framework of Ref.~\cite{Ilten:2018crw}. The exclusion region from an ATLAS search for dilepton resonances~\cite{Aad:2019fac} is also shown in light gray, using the software developed in Ref.~\cite{Kahlhoefer:2019vhz}.}
    \label{fig:exclusionRegion}
\end{figure}

Extra $U(1)$ gauge symmetries are common extensions of the SM~\cite{Davidson:1978pm,Marshak:1979fm,Mohapatra:1980qe,Khalil:2006yi}. Many phenomenological studies sensitive to both heavy and light $Z^\prime$ mediators have been completed combining beam dump experiments and direct searches in colliders~\cite{Erler:1999ub,Langacker:2008yv,Salvioni:2009jp}, and even to explain the anomalous magnetic moment of the muon~\cite{Gninenko:2001hx,Baek:2001kca,Pospelov:2008zw}. In this work, a gauged $B-L$ symmetry is studied, namely that the extra gauge boson couples to quarks and leptons. In this scenario, quarks have $U(1)_{B-L}$ charge $Q_{q}=1/3$, while leptons have  $Q_l=-1$\footnote{These constraints are similar to scenarios of gauged $B-3L_e$ \cite{Denton:2018xmq,Heeck:2018nzc}, $B-2L_e-L_{\mu,\tau}$ and $B-L_e-2L_{\mu,\tau}$ \cite{Flores:2020lji}.}. This will induce the following Beyond the SM interaction between neutrinos and quarks: 
\begin{equation}
    \mathcal{L}_\mathrm{eff} = -\frac{g'^2 Q_l Q_q}{q^2 + M_{Z'}^2}\left[ \sum_\alpha  \bar{\nu}_\alpha \gamma^\mu P_L \nu_\alpha \right] \left[ \sum_q  \bar{q}\gamma_\mu q \right],
\end{equation}
where $q$ is the transferred momentum. This interaction will give rise to interference with the SM cross-section.\\
\indent In Fig.~\ref{fig:exclusionRegion} the expected sensitivities from the detectors are shown for all setups in the $g'-M_{Z'}$ plane. The limits for a one-year exposure are better than other current CE$\nu$NS experiments for all setups. The scintillating bubble chamber would be the leading technology in new vector boson searches from $20$ MeV to ${\sim} 1$~GeV and from $70$ to $230~$GeV.

\subsection{The Neutrino Magnetic Moment}
Neutrino magnetic moments can arise from their interaction with the electromagnetic field, either for Majorana or Dirac neutrinos~\cite{Kayser:1981nw,Vogel:1989iv}. This new interaction contributes to the CE$\nu$NS cross-section without interference, with the following expression:
\begin{equation}
    \frac{d\sigma}{dT} = \pi\frac{\alpha_{\mathrm{EM}}^2Z^2 \mu_\nu^2}{m_e^2} \left(\frac{1}{T} - \frac{1}{E_\nu} + \frac{T}{4E_\nu^2}  \right) F^2(q^2),
\end{equation}
where $\alpha_{\mathrm{EM}}$ is the electromagnetic coupling and $m_e$ is the electron mass. The neutrino magnetic moment, $\mu_{\nu},$ is normalized by the Bohr magneton $\mu_B$.\\
\begin{figure}[t]
    \centering
    \includegraphics[width=\linewidth]{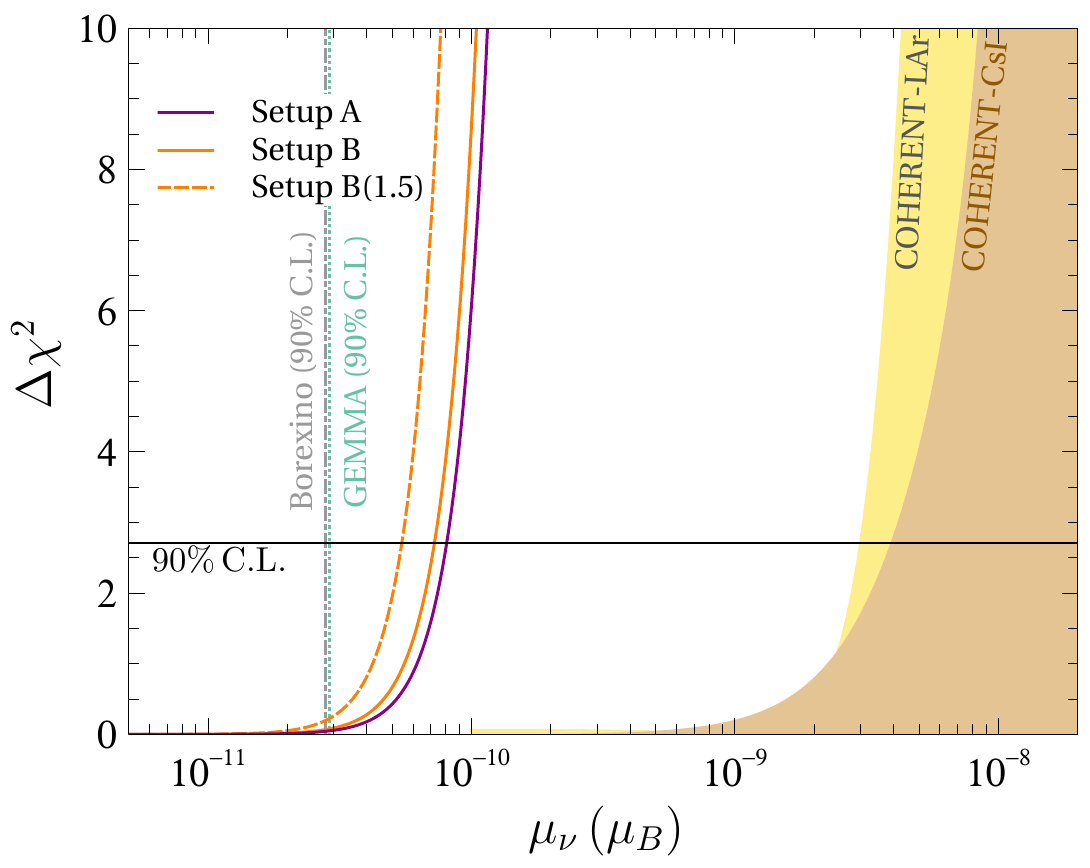}
    \caption{Limits for the neutrino magnetic moment. The solid purple, solid orange and dashed orange lines represent the limits for setups A, B and B(1.5), respectively. The shaded brown and yellow regions correspond to the exclusions set by COHERENT, using CsI~\cite{Akimov:2017ade} and LAr~\cite{Akimov:2020pdx,Akimov:2020czh} detectors, respectively. The GEMMA~\cite{Beda_2013} and Borexino~\cite{Borexino:2017fbd} values are 90\% C.L. bounds.}
    \label{fig:magneticMoment}
\end{figure}
\indent The resulting limits from the $\chi^2$ analysis for the three setups are presented in Fig.~\ref{fig:magneticMoment}. The bounds on the neutrino magnetic moment are of the same order of magnitude as the current GEMMA ($2.9 \times 10^{-11}\mu_B$ at $90\%$ C.L.)~\cite{Beda_2013} and Borexino ($2.8 \times 10^{-11}\mu_B$ at $90\%$ C.L.)~\cite{Borexino:2017fbd} bounds. 

\section{Conclusions}

  The physics reach of a low threshold LAr scintillating bubble chamber for CE$\nu$NS in a reactor has been investigated. A Monte Carlo simulation has shown that it is possible to reach a background level approximately $10\%$ of the signal ({\it in-situ} measurements would constrain the associated systematic uncertainties).
  A plan to determine the nuclear recoil efficiency at a 100~eV energy threshold has been evaluated with the Monte Carlo model developed, showing that it is possible to calibrate to sub-keV energy thresholds using photo-neutron and Thomson scattering sources.
  The sensitivity for an electroweak precision test, a new vector mediator, and the neutrino magnetic moment is very competitive under realistic assumptions for backgrounds and systematic uncertainties. A precision as good as $1\%$ is obtained in the case of the weak mixing angle, a value of the same order as the uncertainty from APV. The setups considered here would set the most stringent bounds for new gauge vector bosons in the $20$ MeV to ${\sim} 1$ GeV and $70$ to $230$ GeV mass ranges. For the neutrino magnetic moment, the best scenario gives a bound of $5.4 \times 10^{-11}\mu_B$ ($90\%$ C.L.), of the same order of magnitude as the current GEMMA ($2.9 \times 10^{-11}\mu_B$ at $90\%$ C.L.) and Borexino ($2.8 \times 10^{-11}\mu_B$ at $90\%$ C.L.) limits. This detector technology has the potential to lead different physics scenarios for coherent elastic neutrino-nucleus scattering experiments and a world leading physics programme can be achieved not only in a power reactor facility (2000 MW$_{th}$), but also in a low power research reactor (1 MW$_{th}$) with only a one-year exposure.
  
\section{Acknowledgements}
\begin{acknowledgments}
\indent The authors  would like to thank Jens Erler for useful comments. This work is supported by the German-Mexican research collaboration grant SP 778/4-1 (DFG) and 278017 (CONACYT), the projects CONACYT CB-2017-2018/A1-S-13051 and CB-2017-2018/A1-S-8960,  DGAPA UNAM grants PAPIIT-IN107621, PAPIIT-IN107118 and PAPIIT-IN108020, and Fundaci\'on Marcos Moshinsky.\\
\indent This work is also supported by FNAL LDRD 2018-003, DOE DE-SC0015910 0003 and DE-SC0011702, and NSF awards 1828609 and 1936432.\\
\indent The authors wish to acknowledge the support of the Natural Sciences and Engineering Research Council of Canada (NSERC) and the Canada Foundation for Innovation (CFI) for funding, Compute Canada (www.computecanada.ca) and the Centre for Advanced Computing, ACENET, Calcul Quebec, Compute Ontario and WestGrid for computational support and the Arthur B. McDonald Canadian Astroparticle Physics Research Institute.\\
\indent The work of LJF is supported by a CONACYT posdoctoral fellowship,
The work of M.~Bressler is supported by a DOE Graduate Instrumentation Research Award. IUSB acknowledges the work of Minji Yun, Marian High School, 1311 South Logan St. Mishawaka, IN 46544.

\end{acknowledgments}
\bibliographystyle{apsrev4-1}
\bibliography{bibliography}
\end{document}